\documentclass[11pt]{article}

\usepackage[xdvi]{graphicx} 
\usepackage{epsfig}
\usepackage{amsfonts,amssymb,amsmath}

\newcommand{\sect}[1]{\setcounter{equation}{0}\section{#1}}
\renewcommand{\theequation}{\arabic{section}.\arabic{equation}}

\newcommand{\ga}{\ensuremath{\gamma}}

\newcommand{\de}{\ensuremath{\delta}}

\newcommand{\ep}{\ensuremath{\epsilon}}

\newcommand{\ka}{\ensuremath{\kappa}}
\newcommand{\la}{\ensuremath{\lambda}}

\newcommand{\om}{\ensuremath{\omega}}

\newcommand{\p}{\ensuremath{\phi}}
\renewcommand{\P}{\ensuremath{\Phi}}
\newcommand{\s}{\ensuremath{\sigma}}

\renewcommand{\th}{\ensuremath{\theta}}
\renewcommand{\d}{\ensuremath{{\rm d}}}
\newcommand{\del}{\ensuremath{\partial}}

\newcommand{\td} {\ensuremath{\tilde}}

\newcommand{\be}{\begin{equation}}
\newcommand{\ee}{\end{equation}}

\newcommand{\ba}{\begin{eqnarray}}
\newcommand{\ea}{\end{eqnarray}}

\begin{document}

\bigskip
\rightline{hep-th/0309058}
\rightline{UUITP-15/03}

\bigskip
\bigskip
\bigskip

\begin{center}
{\Large \bf Rotating Black Holes in a G\"odel Universe}
\end{center}

\bigskip
\bigskip
\bigskip

\centerline{\bf Dominic Brecher$^\flat$, Ulf H. Danielsson$^\natural$,
 James P. Gregory$^\natural$ and Martin E. Olsson$^\natural$}

\bigskip
\bigskip
\bigskip

\centerline{\it $^\flat$Department of Physics and Astronomy}
\centerline{\it University of British Columbia}
\centerline{\it Vancouver, British Columbia V6T 1Z1, Canada}
\centerline{\small \tt brecher@physics.ubc.ca}

\centerline{$\phantom{and}$}

\centerline{\it $^\natural$Department of Theoretical Physics}
\centerline{\it Uppsala University, Box 803, SE-751 08 Uppsala,
  Sweden}
\centerline{\small \tt Ulf.Danielsson, James.Gregory,
  Martin.Olsson@teorfys.uu.se}

 

\bigskip
\bigskip

\begin{abstract}
\vskip 2pt We construct a five--dimensional, asymptotically G\"{o}del,
 three--charge black hole \emph{via} dimensional reduction of an
asymptotically plane wave, rotating D1-D5-brane solution of type IIB
supergravity.  This latter is itself constructed \emph{via} the
solution generating procedure of Garfinkle and Vachaspati, applied to
the standard rotating D1-D5-brane solution.  Taking all charges to be
equal gives a ``BMPV G\"{o}del black hole'', which is closely related
to that recently found by Herdeiro.  We emphasise, however, the
importance of our ten--dimensional microscopic description in terms of
branes.  We discuss various properties of the asymptotically G\"{o}del
black hole, including the physical bound on the rotation of the hole,
the existence of closed timelike curves, and possible holographic
protection of chronology.
\end{abstract}

\newpage

\baselineskip=18pt
\setcounter{footnote}{0}

\sect{Introduction}

There are many conceptual questions in general relativity which are
not expected to be answered within the classical theory itself.  A
typical example is that of classical solutions with closed timelike
curves (CTCs), perhaps the most famous of which is the four--dimensional
G\"odel universe~\cite{Goedel:original} and various generalisations of
it.  Such solutions have CTCs for all times, so it is unclear as to what extent
the arguments of Hawking~\cite{hawking:92} concerning chronology protection ---
themselves quantum mechanical in nature --- can be applied.

Much of the recent interest in such G\"odel--like solutions is due to
the work of~\cite{gauntlett}, where it was shown that
supersymmetric generalisations of the four--dimensional G\"odel
universe are solutions of minimal supergravity in five dimensions.
Moreover, such solutions were further lifted to M--theory in a simple
manner.  Subsequent
work~\cite{Boyda:goedel} made the remarkable
observation that  metrics of the G\"odel type are T--dual to plane
waves.  It was demonstrated there that a wide class of type IIA
G\"odel--like solutions could be found by considering the T--duals of
various
type IIB plane wave solutions.  This was later
applied in~\cite{Harmark:goedel} to produce an array of supersymmetric
G\"odel universes in string theory.

One would expect, therefore, that any problems arising from CTCs on the G\"odel
side of the T--duality would have a mirror on the plane wave side.  It
is important to note, however, that problems on the plane wave
side appear only after the compactification involved in the T--duality
operation.  After all, plane waves are well--defined string
backgrounds with no apparent conceptual problems.  CTCs appear in
\emph{compact} plane waves, however~\cite{brecher:03}, and in quite
some generality~\cite{hubeny:03}.  Whilst one might hope that string
theory somehow resolves any conceptual difficulties related to CTCs,
there is evidence to suggest that string propagation on such compact
plane waves --- and so also on G\"odel backgrounds --- is
problematic~\cite{brace:03}, although the issues are somewhat subtle.

Another example of metrics with CTCs which has been analysed within the
framework of string theory, is the so called BMPV black
hole~\cite{Breckenridge:BMPV} (see also~\cite{Gauntlett:BMPV}).  Here,
however, the issues are well understood.  There is a bound on the angular momentum of
the black hole which, if satisfied, gives a solution with CTCs, but which
are entirely hidden behind the horizon.  Only in the over--rotating
case do the CTCs appear outside of the would--be horizon, although the
solution in that case does not describe a black hole as
such~\cite{gibbons:99}.  These latter backgrounds, which potentially
have CTCs throughout the spacetime, have indeed been
shown to be unphysical for various
reasons~\cite{gibbons:99,Herdeiro:rotate,Jarv:CTC,Dyson:chronology},
and can therefore be safely discarded.  Whether the CTCs in G\"odel--like
backgrounds can be resolved in such ways is, however, an important
unanswered question.  In~\cite{Boyda:goedel} it was suggested
that holography might play an important role, in particular that
holographic screens might shield the CTCs, although the validity of this
proposal is under debate~\cite{Hikida:goedel,brecher:03}.  Other
attempts to exclude G\"odel--like backgrounds on physical grounds
include~\cite{drukker:03}.  In the sense that the CTCs associated with the
G\"{o}del background could possibly be
removed or avoided they are not as pathological as those associated
with the over--rotating black hole.

The aim of this paper is to generalise the (three--charge) BMPV black
hole to a spacetime which is asymptotic to the five--dimensional G\"odel
universe, yet retains the horizon.  In agreement with earlier
results~\cite{Herdeiro:rotate,Gimon:ppblackstrings,Herdeiro:BMPVgoedel}
we show that the G\"odel deformation of the black hole becomes
irrelevant close to the horizon.  The horizon area and entropy
is unchanged from the asymptotically flat case.  Whereas such a
solution has been found already in~\cite{Herdeiro:BMPVgoedel}, we work
within the context of dimensional reduction from ten dimensions, so
obtain from the outset an explicit ten--dimensional microscopic
description of this ``BMPV G\"odel black hole''.  Such a description
would be hard to find from the solution of~\cite{Herdeiro:BMPVgoedel}
directly and, indeed, we show that our microscopic understanding of
this five--dimensional solution has various important implications.
Moreover, although we will not analyse it in much detail, we actually
find a more general BMPV G\"odel black hole to that considered
in~\cite{Herdeiro:BMPVgoedel} and, again, it should be emphasised that
this comes about only through the ten--dimensional description.  We should also note that all the
solutions we consider will be extremal.  We leave the non--extremal
generalisation of our solution for future work; it would be
interesting to see if applying our method to this case
gives the non--extremal solution of~\cite{Herdeiro:BMPVgoedel}.

Our asymptotically G\"odel black hole allows, amongst other things,
for a more detailed investigation of holographic screens and CTCs. In
particular, it is shown that as long as one stays below an upper bound
on the angular momentum for the black hole, there will always be a
causally safe region inside the holographic screen.  Moreover, only
when one violates this bound on the angular momentum, thereby
creating CTCs just outside the black hole horizon, can such
backgrounds exhibit causally sick behaviour \emph{everywhere}.

The CTCs associated with the black hole, and those associated with the
G\"odel background, are of a very different nature.  We hope that our
solution, which combines both types of CTC, will be useful in shedding
light on how, if at all, the CTCs in the G\"odel universe can be
resolved.

In the following section, we generate a solution of type IIB
supergravity which describes rotating D1-D5-branes in a plane wave
background\footnote{Other solutions describing (intersecting) D-branes in plane wave
  backgrounds have been considered in~\cite{bain:02,biswas:02,ohta:03}.}.  As
in the non--rotating case considered
in~\cite{Liu:horizons}, this is achieved by applying the
solution--generating
method~\cite{Garfinkle:v1,Garfinkle:v2,Garfinkle:v3} of Garfinkle and
Vachaspati (GV) to the standard solution describing rotating
D1-D5-branes.  This technique has been known for some time, though has
most often been used to generate asymptotically pp--wave solutions.
To the best of our knowledge, the first time it was used to generate
asymptotically plane wave solutions was in~\cite{hubeny:02}.  In
section three, we T--dualise this to generate a
type IIA solution describing rotating branes in a G\"odel background,
many of the properties of which are inherited by our five--dimensional
G\"odel black hole.  We construct this in section four by dimensional
reduction of the ten--dimensional solution.  In section five, we
analyse the properties of this black
hole, discussing its horizon area, the associated bound on the angular
momentum of the hole, the existence of CTCs and the possible
holographic protection of chronology.  We conclude in section six.

We include two appendices describing our conventions.  The first lists the
field equations of the ten--dimensional type IIB and IIA supergravity
theories, as well as the relevant T--duality rules which map between
them.  The second describes the dimensional reduction of the IIB
theory to five dimensions, generalising somewhat the analysis
of~\cite{herdeiro:02}.


\sect{Asymptotically plane wave rotating branes}

We begin with a type IIB solution representing intersecting
D1-D5-branes with non--trivial angular momentum\footnote{The
Brinkmann, or pp--wave, term discussed in~\cite{Herdeiro:rotate} is
not included here as it will be
generated alongside the plane wave term in the subsequent analysis.
Relative to the solution found in~\cite{Herdeiro:rotate}, we have
absorbed a sign into $C_2$ and switched $\phi_1 \leftrightarrow
\phi_2$, so that the polar parametrisation of $\mathbb{R}^4$ is given
by $z_1+i z_2 = r\cos\th\,e^{i\phi_1}, z_3+i z_4 =
r\sin\th\,e^{i\phi_2}$.}~\cite{Herdeiro:rotate}:
\begin{eqnarray}
\d s^2 & = & H_1^{-\frac{3}{4}}H_5^{-\frac{1}{4}}\left[-\d t^2 + \d y_5^2 +
  \frac{J}{r^2}\left(\cos^2\th\,\d\phi_1-\eta\sin^2\th\,\d\phi_2\right)
  \left(\d t -\d y_5\right)\right] \nonumber \\
  & & \quad
  + H_1^{\frac{1}{4}}H_5^{ \frac{3}{4}}\d s^2(\mathbb{R}^4)
  + H_1^{\frac{1}{4}}H_5^{-\frac{1}{4}}\d s^2(T^4),
  \label{eqn:seedmetric} \\
C_2 & = & \left(H_1^{-1}-1\right) \d t \wedge \d y_5 -
  \frac{J}{2r^2} H_1^{-1} (\d t - \d y_5) \wedge
  \left(\cos^2\th\,\d\phi_1-\eta\sin^2\th\,\d\phi_2\right) \nonumber\\
&& \qquad - \eta Q_5 \cos^2\th\,\d\phi_1\wedge\d\phi_2,
  \label{eqn:C2}  \\
e^{2\Phi} & = & H_1 H_5^{-1}, \label{eqn:Phi}
\end{eqnarray}
where, in terms of the
coordinates $y_1, \ldots, y_4$, $\d s^2(T^4)$ is the flat metric on
$T^4$, the metric on $\mathbb{R}^4$ is
\begin{equation}
\d s^2(\mathbb{R}^4) = \d r^2 + r^2\left(\d\th^2 +
\cos^2\th\,\d\phi_1^2 + \sin^2\th\,\d\phi_2^2\right) \equiv
\sum_{i=1}^4\d z_i^2,
\end{equation}
and the functions associated with the D1- and D5-branes are
\begin{equation}
H_1 = 1 + \frac{Q_1}{r^2}, \qquad H_5 = 1 + \frac{Q_5}{r^2}.
\end{equation}
We have further introduced an arbitrary parameter, $\eta = \pm 1$, into the
solution.  This allows one to
consider either D5-branes ($\eta=+1$ and positive charge), or
anti--D5-branes ($\eta=-1$ and negative
charge).  As noted in~\cite{Herdeiro:BMPVgoedel}, the $SO(4) \simeq
SU(2)_L \times SU(2)_R$ symmetry of the metric (\ref{eqn:seedmetric}),
allows one to define a left, $J_L$, and a right, $J_R$, angular
momentum.  They are just the Casimir operators of $SU(2)_L$ and
$SU(2)_R$. Then, since the rotation one--form,
\be
\cos^2 \th \,\d \p_1 - \eta \sin^2 \th \,\d \p_2,
\ee
appearing in the metric (\ref{eqn:seedmetric}) is a right (left)
one--form of $SU(2)$ for $\eta=+1$ ($\eta=-1$), the solution with
$\eta = +1$ has non--zero $J_R$, whereas that with $\eta = -1$ has
non--zero $J_L$.

To generate the asymptotically plane wave generalisation of
this seed solution, we work with lightcone
coordinates
\begin{equation}
u = \frac{1}{2} (t - y_5), \qquad v = \frac{1}{2} (t + y_5),
\end{equation}
so that $\del/\del u$ is a null Killing vector.  We can then exploit a
fact demonstrated by GV: the presence of
such a Killing vector leads to the separation of the Ricci tensor into
a standard and a plane wave
component~\cite{Garfinkle:v1,Garfinkle:v2,Garfinkle:v3}.  The
appropriate method is discussed by Liu {\it et al}~\cite{Liu:horizons}
who, amongst other things, considered the non--rotating D1-D5-brane
solution, showing that the $r=0$ horizon of the
seed solution is preserved by the GV procedure.  However, it is not
clear from the analysis of~\cite{Liu:horizons} that the horizon is
\emph{regular}\,\footnote{We thank Mukund Rangamani and Nobuyoshi Ohta
for a discussion of this point.}.  Indeed we would expect, as in the
simpler examples considered in~\cite{brecher:00}, that the deformation
we introduce actually destabilises the would--be horizon, giving rise
to a pp--curvature singularity at $r=0$.  In the non--rotating case,
this has been confirmed in~\cite{ohta:03}, and it seems unlikely that
the addition of rotation changes this behaviour.  Of course, the process of T--duality and/or
dimensional reduction to give an asymptotically G\"{o}del solution, would
``remove'' such singular behaviour, giving rise to a perfectly regular
horizon at $r=0$.  Note that 
there are certain types of plane wave deformations which do \emph{not}
give rise to such singularities~\cite{hubeny:02}, but these are somewhat more
complicated than those considered here.  At any rate, relative to the
analysis of~\cite{Liu:horizons}, we will find that we
can add rotation for free.  

As in~\cite{Liu:horizons}, we write
\begin{eqnarray}
\d s^2 & = & H_1^{-\frac{3}{4}}H_5^{-\frac {1}{4}}\left[-4\d u \,\d v +
  \mathcal{H}\d u^2 +
  \frac{2J}{r^2}\left(\cos^2\th\,\d\phi_1-\eta\sin^2\th\,\d\phi_2\right)
  \d u \right] \nonumber \\
  & & \quad
  + H_1^{\frac{1}{4}}H_5^{ \frac{3}{4}}\d s^2(\mathbb{R}^4)
  + H_1^{\frac{1}{4}}H_5^{-\frac{1}{4}}\d s^2(T^4).
\end{eqnarray}
where, asymptotically, we want $\mathcal{H}$ to be a quadratic
function of the transverse coordinates,
\begin{equation}
\mathcal{H} = \sum_{i,j=1}^4\left(A_{ij}y_i y_j + B_{ij}z_i z_j\right).
\end{equation}
The behaviour of $\mathcal{H}$ will be further restricted by the
analysis which follows.  The relevant equations of motion are as in
(\ref{eqn:iib}).  We
must support the plane wave term with an
appropriate contribution to the stress--energy tensor so that the
Einstein equations for our solution are still satisfied.  Since the
Ricci tensor separates, the plane
wave term affects only the $uu$--component of the modified Ricci
tensor, $\mathcal{R}_{ab}$.  The natural
candidate to support this modification of the Einstein
equations is a null five--form\footnote{There are obviously other
possibilities, though we will not consider them here.}.

The five--form must also be self--dual and satisfy $F_5 \wedge F_3 =
0$.  Two natural choices are~\cite{Liu:horizons,Harmark:goedel}
\begin{eqnarray}
F_5 & = & \mu\,\d u \wedge \left( \d y_1 \wedge \d y_2 \wedge \d z_1
\wedge \d z_2 + \d y_3 \wedge \d y_4 \wedge \d z_3 \wedge \d z_4
\right), \label{eqn:F5_Liu} \\
F_5 & = & \frac{\mu}{\sqrt{2}}\,\d u \wedge \left(\d y_1 \wedge \d y_2
+ \d y_3 \wedge \d y_4\right)\wedge\left(\d z_1 \wedge \d z_2 + \d z_3
\wedge \d z_4\right). \label{eqn:F5_Harmark}
\end{eqnarray}

\noindent The dilaton and $F_3$ equations are unchanged, and all that
remains to be checked is the
Einstein equation.  The Ricci tensor of our asymptotically plane wave metric is
given in terms of the unperturbed Ricci tensor,
$\overline{\mathcal{R}}_{ab}$, by
\begin{equation}
\mathcal{R}_{ab} = \overline{\mathcal{R}}_{ab} - \de_a^u \de_b^u
\frac{1}{2 H_1 H_5}\left(\Box + H_5\widehat{\Box} +
\frac{1}{4}(h_5^2+3\,h_1^2)\right)\mathcal{H},
\end{equation}
where
\begin{equation}
h_1=\frac{1}{H_1}\,\frac{\del}{\del r}H_1, \qquad
h_5=\frac{1}{H_5}\,\frac{\del}{\del r}H_5,
\end{equation}
and $\Box$, $\widehat{\Box}$ are the Laplacian operators on
$\mathbb{R}^4$, $T^4$ respectively.  The normalisation of
$F_5$ in (\ref{eqn:F5_Liu}) and (\ref{eqn:F5_Harmark}) has been
chosen in such a way that, for both,
\begin{equation}
F_{a c_1 \ldots c_4} F_b^{~c_1 \ldots c_4} = \frac{48 \mu^2}{H_1 H_5}
\de_a^u \de_b^u.
\end{equation}
The $\mathcal{H}$--deformed
components of the terms arising from $F_3$ are
\be
F_{u ab}F_u^{~ab}  =  \frac{16 J^2}{r^8
  H_1^\frac{5}{2}H_5^\frac{3}{2}} -
  \frac{2h_1^2}{H_1^\frac{3}{2}H_5^\frac{1}{2}}\mathcal{H}, \qquad
  g_{uu} F_3^2 =  
  \frac{6\left(h_5^2-h_1^2\right)}{H_1^\frac{3}{2}H_5^\frac{1}{2}}
  \mathcal{H},
\ee
and one can thus show that the $uu$--component of the Einstein equations is
satisfied provided
\begin{equation}
\left(\Box + H_5\widehat{\Box}\right)\mathcal{H} = -\mu^2.
\end{equation}
Despite the addition of angular momentum, this
equation is unchanged from that derived by Liu
{\it et al}~\cite{Liu:horizons} in the non--rotating case.

The general spherically symmetric solution to this Poisson equation is
a function of the radii, $r$ and $\hat{r}$, of $\mathbb{R}^4$ and
$T^4$ respectively.  One can thus find a general solution which
asymptotes to the maximally supersymmetric BFHP plane
wave~\cite{Blau:planewave},
\begin{equation}
\mathcal{H}=\frac{4Q_k}{r^2} -
\frac{\mu^2}{16}\left(r^2+\hat{r}^2-4Q_5\ln r\right),
\end{equation}
just as in~\cite{Liu:horizons}.  Our objective is to
compactify on the four--torus, however, so we
only consider a solution $\mathcal{H} = \mathcal{H}(r)$, taking
\begin{equation}
\mathcal{H} = \frac{4Q_k}{r^2} - \frac{\mu^2}{8}r^2.
\end{equation}
One could clearly also include an additive constant in the solution
for $\mathcal{H}$, but this may subsequently be absorbed by shifts in
$u$.

Our asymptotically plane wave solution is thus
\begin{eqnarray}
\d s^2 & = & H_1^{-\frac{3}{4}}H_5^{-\frac {1}{4}}\left[-4\d u \,\d v +
  \left(\frac{4Q_k}{r^2} - \frac{\mu^2}{8}r^2\right)\d u^2 +
  \frac{2J}{r^2}\left(\cos^2\th\,\d\phi_1-\eta\sin^2\th\,\d\phi_2\right)
  \d u \right] \nonumber \\
  & & \quad
  + H_1^{\frac{1}{4}}H_5^{ \frac{3}{4}}\d s^2(\mathbb{R}^4)
  + H_1^{\frac{1}{4}}H_5^{-\frac{1}{4}}\d s^2(T^4),
\end{eqnarray}
supported by the dilaton, (\ref{eqn:Phi}), the RR two--form potential,
(\ref{eqn:C2}), and an RR four--form
potential $C_4$ giving rise to one of the five--forms, $F_5 = \d C_4$,
in (\ref{eqn:F5_Liu}) or (\ref{eqn:F5_Harmark}).  Switching off the
five--form field strength by setting
$\mu=0$, we obtain the rotating D1-D5-pp--wave system given
explicitly in~\cite{Herdeiro:rotate}.  This is the
ten--dimensional description of the familiar five--dimensional
rotating three--charge black
hole~\cite{Cvetic:3charge,Horowitz:3charge}, which is obtained upon
dimensional reduction along the five directions $y_1, \ldots y_5$.
Setting all three charges to be equal, $Q_1=Q_5=Q_k$, we recover
the BMPV black hole~\cite{Breckenridge:BMPV}.

If instead we switch off the black hole charges,
setting $Q_1 = Q_5 = Q_k = J = 0$, then we obtain a standard supersymmetric
plane wave as expected.  The
observation~\cite{Boyda:goedel,Harmark:goedel} that this is T--dual to
a type
IIA G\"odel universe, $\mathcal{G}_5 \times
\mathbb{R}^5$,
implies that the solution presented here will be T--dual
to a rotating D0-D4-F1 system in a G\"odel universe.  We will demonstrate
that this is indeed the case in the next section.


\sect{Rotating branes in a rotating universe}

\subsection{Constructing the IIA solution}

In order to T--dualise this solution, we must choose an
everywhere spacelike Killing vector, $K$, along which to perform the
T--duality.  To generate a G\"{o}del--like universe one takes $K$ to
be a translation in $y_5$ plus rotations in the transverse
planes~\cite{Boyda:goedel,Harmark:goedel}.  For the plane wave, the
quotient by the action
of such a $K$ leads to CTCs~\cite{brecher:03} which are inherited by
the G\"{o}del solution.  More generally, the
quotient of a plane wave by the action of any Killing vector with a
$\del /\del u$ component will give rise to a G\"{o}del--like universe
with CTCs~\cite{hubeny:03}, and we expect the same to be true here.
We thus take
\begin{equation}
K = \frac{\del}{\del y_5} - \left(\beta_1 \frac{\del}{\del\phi_1} + \beta_2
\frac{\del}{\del\phi_2}\right),
\label{eqn:k}
\end{equation}
which has norm
\begin{equation}
\vert K \vert^2 = 1 + \zeta(r,\th) \equiv H_{k\beta}(r,\theta),
\label{eqn:H}
\end{equation}
where
\begin{equation}
\zeta(r,\th) = \left(H_1 H_5(\beta_1^2\cos^2\th+\beta_2^2\sin^2\th) -
\frac{\mu^2}{32}\right)r^2 + \frac{Q_k}{r^2}+\frac{J}{r^2}
\left(\beta_1\cos^2\th-\eta\beta_2\sin^2\th\right).
\end{equation}
The simplest way to ensure $|K|^2 > 0$ everywhere is to
take $\beta_1^2=\beta_2^2=\mu^2/32 =
\beta^2$.  Defining $\beta_1 = \beta$, $\beta_2 = \ep\beta$, where $\ep=\pm 1$,
we have
\begin{equation}
\zeta(r,\th) = \beta^2(Q_1+Q_5)+\frac{Q_{k\beta}(\th)}{r^2},
\end{equation}
where
\begin{equation}
Q_{k\beta}(\th) = Q_k +\beta^2Q_1Q_5+\beta J
\left(\cos^2\th - \ep\eta\sin^2\th\right).
\label{eqn:Qkb}
\end{equation}

With both $\beta$ and $J$ positive, and with the
choice $\ep = -\eta$, no further restriction is required.  In this
case, the twist
induced by $K$ has the same sign relative to the original angular
momentum, governed by $J$, in each of the two planes of rotation.
With $\ep = +\eta$, however, this is not the case, and we need to impose
a further restriction,
\begin{equation}
Q_k +\beta^2 Q_1 Q_5 \ge \beta J,
\label{eqn:restrict}
\end{equation}
on the charges to ensure that $K$ is everywhere spacelike.  (Note that
if we instead take either $\beta$ or $J$ to be negative, then the same
condition should also be imposed.)  This
choice further introduces a $\th$ dependence in the metric.  Either way,
we can now T--dualise along the orbits of $K$.  For simplicity, when studying
solutions which contain both background G\"odel rotation, $\beta$, and
rotation, $J$, of the branes, we
will usually restrict to the case $\ep = -\eta$ (although see the
discussion of minimal supergravity in the following section).  This
corresponds to
taking the two types of rotations, induced by $\beta$ and $J$,
to contribute to the same angular momentum, $J_R$ ($J_L$) for
$\eta=+1$ ($-1$) respectively.

Introducing a new pair of angular coordinates
\begin{equation}
\widetilde{\phi}_1 = \phi_1 - 2\beta_1 u, \qquad
\widetilde{\phi}_2 = \phi_2 - 2\beta_2 u,
\end{equation}
which satisfy $K(\widetilde{\phi_i}) = 0$, the Killing vector along
which we T--dualise becomes $\del/\del y_5$.  With $\d s^2
(\widetilde{\mathbb{R}}^4)$ denoting the metric on $\mathbb{R}^4$ in
terms of these new coordinates, the resulting metric is
\begin{eqnarray} \label{eqn:twistmetric}
\d s^2 & = & H_1^{-\frac{3}{4}}H_5^{-\frac{1}{4}}H_{k\beta} \left(\d y_5 -
H_{k\beta}^{-1}(\zeta(r,\th) \d t + \s)\right)^2 + H_1^{\frac{1}{4}}
H_5^{\frac{3}{4}}
\d s^2(\widetilde{\mathbb{R}}^4) \nonumber \\
&& \quad 
-H_1^{-\frac{3}{4}} H_5^{-\frac{1}{4}}H_{k\beta}^{-1} \left( \d t - \s
\right)^2 + H_1^{\frac{1}{4}}H_5^{-\frac{1}{4}}
\d s^2(T^4),
\end{eqnarray}
where we have returned to the coordinates $(t, y_5)$.  The metric has
been written in a way
adapted to dimensional reduction along $y_5$, and we
will utilise this in the following section.  Here we T--dualise
according to the equations (\ref{eqn:tdual}).  We have also defined
the one--form 
\begin{equation} \label{eqn:sigma1}
\s = r^2\left(\ga_1(r)\cos^2\th\,\d\widetilde\phi_1 +
\ga_2(r)\sin^2\th\,\d\widetilde\phi_2\right),
\end{equation}
in which
\be
\ga_1(r) =  \beta_1 H_1H_5 + \frac{J}{2r^4}, \qquad \ga_2(r)  =
\beta_2 H_1H_5 - \frac{\eta J}{2r^4}.
\ee
The RR two--form potential becomes
\be
C_2 =  \left(\left(H_1^{-1}-1\right) \d t + \om\right) \wedge
  \d y_5 + \d t \wedge \om - \eta\,Q_5 \cos^2\th\,\d\widetilde{\phi}_1
  \wedge\d\widetilde{\phi}_2,
\label{eqn:c2}
\ee
where we define the one--form
\begin{equation}
\om = f_2(r,\th)\d\widetilde\phi_1 - \eta f_1(r,\th)\d\widetilde\phi_2,
\end{equation}
in which
\be
f_1(r,\th) =  \phantom{\eta}\beta_1 Q_5 \cos^2\th - \frac{J}{2r^2}
H_1^{-1} \sin^2\th, \qquad f_2(r,\th) =  \left( \eta \beta_2 Q_5  -
\frac{J}{2r^2} H_1^{-1} \right)\cos^2\th.
\ee

With the five--form field strength given by either (\ref{eqn:F5_Liu})
or (\ref{eqn:F5_Harmark}), there is a gauge freedom in choosing a
four--form such that $F_5 = \d C_4$.  Concentrating on the five--form
(\ref{eqn:F5_Liu}) for definiteness, in many ways the simplest
choice is to take
\begin{equation}
C_4 = \frac{\mu(t-y_5)}{2} \left(\d y_1 \wedge \d y_2 \wedge \d
\td{z}_1 \wedge \d \td{z}_2 + \d y_3 \wedge \d y_4 \wedge \d \td{z}_3
\wedge \d \td{z}_4
\right),
\label{eqn:c4}
\end{equation}
which does not contribute to the three--form RR
potential of the IIA solution, giving a five--form alone.  On
the other hand, we could choose
\begin{equation}
C_4 = - \frac{\mu}{2}\left(\d t -\d y_5\right) \wedge \left(y_1\, \d
y_2 \wedge \d \td{z}_1 \wedge \d \td{z}_2 + y_3\, \d y_4 \wedge \d
\td{z}_3 \wedge \d \td{z}_4 \right),
\end{equation}
which gives rise to both three--form and five--form RR potentials in the IIA
solution.

With $C_4$ as in (\ref{eqn:c4}), the resulting IIA solution is
\begin{eqnarray}
\d s^2 & = & -H_1^{-\frac{7}{8}} H_5^{-\frac{3}{8}}
H_{k\beta}^{-\frac{3}{4}} \left(\d t - \s\right)^2 
  + H_1^{\frac{1}{8}}H_5^{\frac{5}{8}}H_{k\beta}^{\frac{1}{4}}
  \d s^2(\widetilde{\mathbb{R}}^4)
  + H_1^{\frac{1}{8}}H_5^{-\frac{3}{8}}H_{k\beta}^{\frac{1}{4}}
  \d s^2(T^4) \nonumber \\
&& \qquad
  + H_1^{\frac{1}{8}}H_5^{\frac{5}{8}}H_{k\beta}^{-\frac{3}{4}}\d
  y_5^2, \nonumber \\
e^{2\Phi} & = & H_1^{\frac{3}{2}}H_5^{-\frac{1}{2}}H_{k\beta}^{-1},
\nonumber \\
B_2 & = & \left(\left(H_{k\beta}^{-1}-1\right)\,\d t - H_{k\beta}^{-1}
  \s\right) \wedge \d y_5, \qquad C_1  =  -\left(H_1^{-1}-1\right)\d t
  - \om, \label{eqn:IIA} \\
C_3 & = & \left(r^2 H_{k\beta}^{-1} \left(\eta\ga_1 f_1\cos^2\th
  +\ga_2 f_2\sin^2\th\right) -\eta Q_5 \cos^2\th\right)
  \d\widetilde{\phi}_1 \wedge \d\widetilde{\phi}_2 \wedge \d y_5
  \nonumber \\
&& \qquad + H_{k\beta}^{-1} \d t \wedge
  \left(\om - Q_1 H_1^{-1}r^{-2}\s \right) \wedge \d y_5, \nonumber \\
C_5 & = & \frac{\mu(t-y_5)}{2}\left(\d y_1 \wedge \d y_2 \wedge \d
  \td{z}_1 \wedge \d \td{z}_2 + \d y_3 \wedge \d y_4 \wedge \d
  \td{z}_3 \wedge \d \td{z}_4
  \right) \wedge \d y_5, \nonumber
\end{eqnarray}
The NS--NS two form, $B_2$,
gives rise to a field strength, $H_3 = \d B_2$, and the RR field
strengths are $F_2 = \d C_1$ and $G_4 =
\d C_3 + C_1 \wedge H_3$.  There is also a six--form field strength,
$G_6 = \d C_5 + C_3 \wedge H_3$, which, as described in appendix A,
can be dualised as $G_4^\prime = -e^{-\Phi/2} \star G_6$ to give a further
contribution to the four--form field strength.  The other choice of five--form,
as in (\ref{eqn:F5_Harmark}), and with a similar choice of gauge as
above, gives rise to an alternative IIA solution with
\begin{equation}
C_5 = \frac{\mu(t-y_5)}{2\sqrt{2}} \left(\d y_1 \wedge \d y_2 + \d
y_3 \wedge \d y_4\right)\wedge\left(\d \td{z}_1 \wedge \d \td{z}_2 +
\d \td{z}_3
\wedge \d \td{z}_4\right) \wedge \d y_5.
\end{equation}

\subsection{Dissecting the IIA solution}

As is easily seen by considering the limits in which more familiar
supergravity solutions arise, the type IIA solution (\ref{eqn:IIA})
corresponds to a rotating
D0-D4-F1 system in a G\"odel universe.  It has CTCs for radii
\be
r^2 \gamma_i^2 \ge H_1H_5H_{k\beta}, \qquad i = 1,2.
\ee
Compactification on $T^4 \times S^1$ will not change
this property of the solution, so we expect it to persist in the
five--dimensional asymptotically G\"{o}del black hole solutions of the
following section.  We do indeed find the same condition there, so
will postpone its analysis until then.

Switching off the G\"{o}del parameter, taking
$\beta = 0$, gives a solution discussed
in~\cite{herdeiro:02,Dyson:chronology}:
\begin{eqnarray}
\d s^2 & = & -H_1^{-\frac{7}{8}} H_5^{-\frac{3}{8}}
H_k^{-\frac{3}{4}} \left(\d t - \s\right)^2 
  + H_1^{\frac{1}{8}}H_5^{\frac{5}{8}}H_k^{\frac{1}{4}}
  \d s^2(\widetilde{\mathbb{R}}^4)
  + H_1^{\frac{1}{8}}H_5^{-\frac{3}{8}}H_k^{\frac{1}{4}}
  \d s^2(T^4) \nonumber \\
&& \quad
  + H_1^{\frac{1}{8}}H_5^{\frac{5}{8}}H_k^{-\frac{3}{4}}\d
  y_5^2, \nonumber \\
e^{2\Phi} & = & H_1^{\frac{3}{2}}H_5^{-\frac{1}{2}}H_k^{-1}, \nonumber\\
B_2 & = & \left( \left( H_k^{-1}-1 \right) \, \d t -
  H_k^{-1} \s \right) \wedge \d y_5, \\
C_1 & = & -\left( H_1^{-1}-1 \right) \d t + H_1^{-1} \s, \nonumber \\
C_3 &=& H_k^{-1} \d t \wedge \d y_5 \wedge \s, \nonumber
\end{eqnarray}
where
\be
\s = \frac{J}{2r^2} \left( \cos^2\th \d \td{\p}_1 - \eta \sin^2\th \d
  \td{\p}_2 \right),
\ee
and
\be
H_k = 1 + \frac{Q_k}{r^2}.
\ee
It describes a rotating D0-D4-F1
system, with (D0,D4,F1) charges $(Q_1,Q_5,Q_k)$.  Compactification on
the $T^4$ would give rise to the five--dimensional rotating three--charge
black hole and, as in that case~\cite{herdeiro:02,Dyson:chronology},
the condition for the existence of CTCs reduces to
\be
J^2 \ge 4r^6H_1H_5H_k.
\ee

The other limiting case is pure G\"{o}del: $Q_1 = Q_5 = Q_k = J = 0$.
In terms of the field strengths $H_3 = \d B_2$ and $F_4 = \star F_6 =
\star (\d C_5)$, the solution is
\ba
\d s^2 &=& - \left( \d t - \s \right)^2 + \d s^2 (\td{\mathbb{R}}^4) +
\d s^2 (T^4) + \d y_5^2, \nonumber \\
H_3 &=& -2 \beta \left( \d \td{z}^1 \wedge \d \td{z}^2 + \ep \d
  \td{z}^3 \wedge \d \td{z}^4 \right) \wedge \d y_5, \\
F_4 &=& 2 \sqrt{2} \beta \left( \d y_1 \wedge \d y_2 \wedge \d \td{z}_1
  \wedge \d \td{z}_2 + \d y_3 \wedge \d
  y_4 \wedge \d \td{z}_3 \wedge \d \td{z}_4 \right),  \nonumber 
\ea
where now
\be
\s = \beta r^2 ( \cos^2\th \d \td{\p}_1 + \ep \sin^2\th \d \td{\p}_2).
\ee
The parity of the rotation in the original Killing vector
(\ref{eqn:k}) thus
determines whether we generate a G\"odel universe with non--zero $J_L$
or $J_R$.  In either case the condition for the existence of
CTCs reduces to  
\be
r^2 > \frac{1}{\beta^2}.
\ee
We will refer to the surface at which such CTCs (associated with the
G\"{o}del rotation) appear as the velocity of light (VL) surface.

The general case has non--zero G\"odel rotation, $\beta$, and ``bare''
brane rotation, $J$.  (Recall that we will set
$\ep=-\eta$, so that both rotations contribute to the single
non--zero angular \
momentum, $J_L$ or $J_R$.)  The metric has the same form as that of a
D0-D4-F1 system, but with a modified one--form, $\sigma$, governing
rotation and a modified harmonic function, $H_{k\beta}$, associated with the
fundamental string.  The one--form is
\be 
\s = \left( \beta r^2 + \beta (Q_1 + Q_5) + \frac{J+2\beta Q_1
  Q_5}{2r^2} \right) (\cos^2\th \d \td{\p}_1-\eta\sin^2\th \d \td{\p}_2),
\ee
in which the ${\cal O}(r^2)$ term is the G\"{o}del rotation, and the ${\cal
  O}(1/r^2)$ term is the rotation of the branes.  It is thus clear
that frame dragging effects associated with the background G\"{o}del
rotation modifies the bare brane rotation to give an effective rotation,
\be \label{eqn:Jbeta}
J_\beta = J + 2\beta Q_1 Q_5.
\ee

Consider now the harmonic function associated with the fundamental string,
\be
H_{k\beta}(r) = \la\left(1 + \frac{\la^{-1}Q_{k\beta}}{r^2}\right),
\ee
where
\be \label{eqn:F1}
\lambda = 1 + \beta^2 (Q_1 +Q_5) \qquad \textrm{and} \qquad Q_{k\beta}
  = Q_k + \beta^2 Q_1 Q_5 + \beta J.
\ee
The G\"odel parameter combines with the bare rotation, and D0- and
D4-brane charges, $Q_1$ and $Q_5$, to give a modified
string charge $\lambda^{-1} Q_{k\beta}$.

At this stage the richness of the ten--dimensional solution is
apparent.  Compactification on $T^4 \times S^1$ gives a five--dimensional
solution with equally rich properties.


\sect{The five--dimensional rotating G\"odel black hole}

By choosing the initial plane wave deformation, $\mathcal{H}$, to be
independent of the transverse coordinates, $y_1,\dots,y_4$, we have
preserved translations in these directions as isometries of our final
solution.  We can thus compactify either the IIB or the IIA solution
on the $T^4\times S^1$ provided by the directions $y_1 \dots y_5$.
Although this will give rise to the same five--dimensional solution,
we follow the type IIB route described in~\cite{herdeiro:02}.  Of
course in that case, we perform a twisted compactification along
$y_5$, so will compactify the rotated
solution~(\ref{eqn:twistmetric}), (\ref{eqn:c2}) and (\ref{eqn:Phi})
on $T^4\times S^1$ in the usual way.

The compactification ansatz derived in~\cite{herdeiro:02} is reviewed
in appendix B.  In addition to the dilaton, which reduces trivially, this
generates a further two scalars, $\psi$ and $\chi$, a gauge field,
$A_1$, coming from the
metric, and two gauge fields, $B_2$ and $C_1$, coming from the RR
two--form.  We find the five--dimensional solution
\ba \label{eqn:GBH}
\d s^2_5 &=& -\left( H_1 H_5 H_{k\beta} \right)^{-\frac{2}{3}} ( \d t
- \s)^2 + \left(
  H_1 H_5 H_{k\beta} \right)^{\frac{1}{3}} \d s^2
  (\widetilde{\mathbb{R}}^4), \nonumber\\ 
A_1 &=& - H_{k\beta}^{-1} \left( \zeta \d t + \s \right), \nonumber \\
B_2 &=& - \eta Q_5 \cos^2\th \d \td{\p}_1 \wedge \d \td{\p} + \d t
\wedge \om, \\
C_1 &=& (H_1^{-1} -1 ) \d t + \om, \nonumber \\
e^{2\P} &=& H_1 H_5^{-1}, \qquad e^{2b\psi} = H_1^{-\frac{1}{3}}
H_5^{\frac{1}{3}},
\qquad e^{2a\chi} = (H_1 H_5)^{-\frac{1}{6}} H_{k\beta}^{\frac{1}{3}},
\nonumber 
\ea
where $a^2=1/24$, $b^2=1/9$ and the one--form, $\s$, is given by
(\ref{eqn:sigma1}).  Note that we have $\P = -3b \psi$, which
is consistent with the equations of motion (\ref{eqn:5d}); there are
effectively only two scalar fields.  The gauge
potentials combine to give the field strengths $F_{(1)} = \d A_1$,
$F_{(2)} = \d C_1$ and $G_3 = \d B_2 - F_{(2)} \wedge A_1$.  The
latter can be dualised as
\be
G_3 = e^{-(\P + 2 ( 2a\chi - 3b\psi/2))} \star \d \hat{A}_1
\equiv e^{-(\P + 2 ( 2a\chi - 3b\psi/2))} \star \hat{F}_2.
\ee
In addition, there are three--form and two--form field strengths coming
from the reduction of the RR five--form.  As described in appendix B,
the choice (\ref{eqn:F5_Liu}) gives rise to
\be
F_3^{ij} = \frac{\mu}{2} \d t \wedge \d
\td{z}_i \wedge \d \td{z}_j, \qquad F_2^{ij} = - \frac{\mu}{2}\d
\td{z}_i \wedge \d \td{z}_j,
\ee
for $i,j=1,2$ and $i,j=3,4$ only.  They combine to give the gauge
invariant field strength
\be
G_3^{ij} = F_3^{ij} - F_2^{ij} \wedge A_1 = \frac{\mu}{2} H_{k\beta}^{-1} ( \d
t - \s ) \wedge \d
\td{z}_i \wedge \d \td{z}_j,
\ee
which can be dualised by taking
\be
G_3^{ij} = e^{-4a \chi} \star \d \hat{A}_1^{ij},
\ee
in terms of the dual potentials
\be
\hat{A}_1^{12} = \frac{\mu}{4}\left( \td{z}_3 \d \td{z}_4 -\td{z}_4
  \d \td{z}_3 \right), \qquad  \hat{A}_1^{34} = \frac{\mu}{4} \left(
  \td{z}_1 \d \td{z}_2 -\td{z}_2  \d \td{z}_1 \right).
\label{eqn:dual}
\ee

On the other hand, the choice (\ref{eqn:F5_Harmark}) gives 
\be
F_3^{12} = F_3^{34} = \frac{\mu}{2\sqrt{2}} \d t \wedge ( \d
\td{z}_1 \wedge \d \td{z}_2 + \d \td{z}_3 \wedge \d \td{z}_4 ), \qquad
F_2^{12} = F_2^{34} = - \frac{\mu}{2\sqrt{2}} ( \d
\td{z}_1 \wedge \d \td{z}_2 + \d \td{z}_3 \wedge \d \td{z}_4 ).
\ee
They combine to give the gauge invariant field strength
\be
G_3^{12} = G_3^{34} = \frac{\mu}{2\sqrt{2}} H_{k\beta}^{-1} ( \d
t - \s ) \wedge ( \d
\td{z}_1 \wedge \d \td{z}_2 + \d \td{z}_3 \wedge \d \td{z}_4 ),
\ee
which can be dualised in terms of
\be
\hat{A}_1^{12} = \hat{A}_1^{34} = \frac{\mu}{4\sqrt{2}}\left( \td{z}_1 \d
  \td{z}_2 -\td{z}_2  \d \td{z}_1 + \td{z}_3 \d \td{z}_4 -\td{z}_4
  \d \td{z}_3 \right).
\label{eqn:dual2}
\ee

We interpret the solution (\ref{eqn:GBH}) as the five--dimensional rotating
three--charge black hole in a G\"odel universe.  Although it is not a
solution of minimal
five--dimensional supergravity, one might expect that the two limits
are.  To recover minimal supergravity, the scalars must vanish.  So
either $\beta=0$ and $Q_1=Q_5=Q_k$ (the BMPV black hole), or
$Q_1=Q_5=Q_k=J=0$ (the G\"{o}del universe).

Setting $\beta=0$ gives the five--dimensional rotating
three--charge black hole~\cite{Cvetic:3charge,Horowitz:3charge}:
\ba \label{eqn:3charge}
\d s_5^2 &=& -(H_1H_5H_k)^{-\frac{2}{3}} \left(\d t - \s \right)^2 +
(H_1H_5H_k)^{\frac{1}{3}}\d s^2(\mathbb{R}^4), \nonumber \\
A_1 &=& \left( H_k^{-1} -1 \right) \d t -H_k^{-1} \s, \nonumber\\
B_2 &=& -\eta Q_5 \cos^2\th \d \td{\p}_1 \wedge \d \td{\p}_2 - H_1^{-1}
\d t \wedge \s, \\
C_1 &=& \left( H_1^{-1} -1 \right) \d t -H_1^{-1} \s, \nonumber\\
e^{2\P} &=& H_1 H_5^{-1}, \qquad e^{2b\psi} = H_1^{-\frac{1}{3}}
H_5^{\frac{1}{3}}, 
\qquad e^{2a\chi} = (H_1 H_5)^{-\frac{1}{6}} H^{\frac{1}{3}}, \nonumber
\ea
where
\be
\s = \frac{J}{2r^2} \left(\cos^2\th\,\d\widetilde\phi_1 -
\eta\sin^2\th\,\d\widetilde\phi_2\right), \qquad H_k = 1
+\frac{Q_k}{r^2},
\ee
and the BMPV black hole~\cite{Breckenridge:BMPV} arises from setting the
charges $Q_1=Q_5=Q_k \equiv Q$.  In this
equal--charge case the scalars do indeed vanish, and we have $F_{(1)}
= F_{(2)} = F$.  To recover the equation
\be
\d \star F + F \wedge F = 0,
\label{eqn:minimal}
\ee
of the minimal theory, we then also need
\be
G_3 = \star F.
\ee
However, an explicit calculation for the solution (\ref{eqn:3charge}) gives
\be
G_3 = + \eta \star F,
\ee
so it would seem that only the right ($\eta=+1$) case is a solution of
minimal supergravity.  In this sense, the minimal theory is the
positive charge sector of ours.

On the other hand, setting the black hole charges to zero gives
\ba
\d s_5^2 &=& -\left(\d t - \s \right)^2 + \d
s^2(\mathbb{R}^4), \nonumber \\
A_1 &=& -\s,
\ea
where now
\be
\s = \beta r^2\left(\cos^2\th\,\d\widetilde\phi_1 +\ep
  \sin^2\th\,\d\widetilde\phi_2\right),
\label{eqn:sigma}
\ee
but there are further dual one--forms (\ref{eqn:dual}) or
(\ref{eqn:dual2}), which came originally from the RR five--form field strength
in ten dimensions.  Since $G_3 = F_{(2)} = 0$, and with $F_{(1)} = \d
A_1 = F$, we would need
\be
G_3^{ij} = c^{ij} \star F, \qquad \sum_{j>i} c^{ij}c^{ij} = 1,
\ee
to get the minimal equation (\ref{eqn:minimal}) as in appendix B.   

It is clear that the potentials given in
(\ref{eqn:dual}) do not fall into this category, although those in
(\ref{eqn:dual2}) do.  The latter will therefore give a solution of the minimal
theory, but only with $\ep=+1$ in (\ref{eqn:sigma}).  The parity of
the Kaluza--Klein vector must match that of the dual one--form coming
from the self--dual five--form in ten dimensions.  (To match with
$\ep=-1$ in (\ref{eqn:sigma}), we would need to start with an
\emph{anti}--self--dual five--form.)  Note that this
implies $\ep = +\eta$, so to get both BMPV and G\"{o}del solutions of
the minimal theory we need to make this choice in (\ref{eqn:Qkb}), and
this leads to a more complicated, $\th$ dependent, metric.  Taking
$\ep=-\eta$ to make the analysis simpler, means that only
our BMPV ($\eta=+1$) \emph{or} our G\"{o}del ($\eta=-1$) solution will
be relevant to the minimal theory.

A different approach was recently used by Herdeiro to construct this
``BMPV G\"odel black hole''~\cite{Herdeiro:BMPVgoedel}, taking the
extremal limit of a Kerr--Newman--G\"{o}del black hole.  This, in turn
was constructed \emph{via} a Hassan--Sen (HS)
transformation~\cite{hassan:92} of a Kerr--G\"{o}del black hole found
in~\cite{gimon:03}.  We can demonstrate that our black hole indeed
recovers the BMPV G\"odel solution of~\cite{Herdeiro:BMPVgoedel} as
follows.  We first consider a rescaling of coordinates by defining
``natural'' time and radial coordinates, $\bar{t}$ and $\bar{r}$, as
those in which the asymptotic behaviour of the general
solution~(\ref{eqn:GBH}) is
\begin{equation}
\d s^2_5 = -\left(\d \bar{t} - \s\right)^2 + \d \bar{r}^2 +\bar{r}^2\d
\Omega^2_3.
\end{equation}
This requires the rescaling
\begin{equation}
\bar{t} = \la^{-1/3} t, \qquad \bar{r} = \la^{1/6} r,
\end{equation}
with $\la$ as in (\ref{eqn:F1}).  Such rescalings necessitate further
rescalings of the dimensionful parameters
\begin{equation}
\bar{Q}_1 = \la^{1/3}Q_1, \qquad
\bar{Q}_5 = \la^{1/3}Q_5, \qquad
\bar{Q}_{k\beta} = \la^{1/3} \left(\la^{-1}Q_{k\beta}\right), \qquad
\bar{\beta} = \la^{-2/3}\beta.
\end{equation}

The black hole metric~(\ref{eqn:GBH}) then takes the form
\be \label{eqn:5D3charge}
\d s_5^2  =  -\mathbb{H}^{-\frac{2}{3}} \left[\d \bar{t} -
  \left(\bar{\beta}\bar{r}^2 + \bar{\beta}(\bar{Q}_1+\bar{Q}_5) +
  \frac{J_\beta}{2\bar{r}^2}\right)
  \left(\cos^2\th\,\d\widetilde\phi_1 -
  \eta\sin^2\th\,\d\widetilde\phi_2\right)\right]^2  +
  \mathbb{H}^{\frac{1}{3}}\,\d \bar{s}^2(\mathbb{R}^4),
\ee
where
\begin{equation} \label{eqn:BMPVgoedel}
\mathbb{H} = \left(1+\frac{\bar{Q}_1}{\bar{r}^2}\right)
\left(1+\frac{\bar{Q}_5}{\bar{r}^2}\right)
\left(1+\frac{\bar{Q}_{k\beta}}{\bar{r}^2}\right).
\end{equation}
Taking $\bar{Q}_1 = \bar{Q}_5 = \bar{Q}_{k\beta} = \bar{Q}$, gives the
BMPV G\"odel black hole of~\cite{Herdeiro:BMPVgoedel}.  To reproduce
this latter precisely, we should identify the parameters of our
solution as
\be
\bar{\beta} \rightarrow 2J, \qquad \bar{Q} \rightarrow \mu, \qquad
J_\beta \rightarrow 4\mu \om,
\ee
giving a ten--dimensional interpretation of the five--dimensional
charges in~\cite{Herdeiro:BMPVgoedel}: the five--dimensional G\"{o}del
parameter, $J$, becomes the ten--dimensional G\"{o}del parameter
$\bar{\beta}$; the five--dimensional mass (or charge), $\mu$, becomes
the ten--dimensional charge $\bar{Q}$; and the five--dimensional
angular momentum, $\om$, is related to the ten--dimensional brane
rotation $J_\beta$.  It is intriguing that the GV procedure in ten
dimensions is equivalent to a HS transformation in six.  We should
emphasise, however, that our understanding of the ten--dimensional
origins of the BMPV G\"odel black hole leads naturally to the more
general solution (\ref{eqn:GBH}), with $H_{k\beta}$ as in
(\ref{eqn:H})--(\ref{eqn:Qkb}).  For $\ep=+\eta$ (the G\"{o}del
rotation and black hole rotation generating opposite angular momenta,
$J_L$ or $J_R$), there is $\th$ dependence in the spacetime fields.
The further restriction (\ref{eqn:restrict}) on the charges must also
be imposed in this case.  It would be interesting to understand this
restriction as arising from the mixed angular momenta solution in the
five--dimensional language, but for now we will concentrate on the
properties of the $\ep=-\eta$ version of the metric~(\ref{eqn:GBH}),
and its rescaled form~(\ref{eqn:5D3charge}).


\sect{Properties of the black hole} \label{sect:properties}

We consider the three--charge black hole in the G\"odel universe given
by the metric~(\ref{eqn:5D3charge}).  As noted
in~\cite{Herdeiro:BMPVgoedel}, for $J^2 < 4
Q_1Q_5Q_k$, the coordinate singularity at
$\bar{r}=0$ is a null surface and can be interpreted as a
horizon\footnote{In the equal charge case, this can be shown
rigorously by introducing
  coordinates which cover the
horizon~\cite{herdeiro:02}.}.  In the BMPV case
($\bar{Q}_1=\bar{Q}_5=\bar{Q}_{k\beta}$), as
in~\cite{Herdeiro:BMPVgoedel} we
can introduce a Schwarzschild--like coordinate
\be
\bar{R}^2=\bar{r}^2+\bar{Q},
\label{eqn:R}
\ee
in which the metric becomes
\be
\d s_5^2  =  -\left( 1 - \frac{\bar{Q}}{\bar{R}^2}\right)^2 \left(\d \bar{t} -
  \s \right)^2  + \left( 1 - \frac{\bar{Q}}{\bar{R}^2}\right)^{-2} \d
\bar{R}^2 + \bar{R}^2 \d \Omega_3^2,
\ee
where
\be
\s = \frac{1}{\bar{R}^2} \left( 1 -
\frac{\bar{Q}}{\bar{R}^2}\right)^{-1} \left( 
  \bar{\beta} ( \bar{R}^4 - \bar{Q}^2) + \frac{J_\beta}{2} \right)
\left(\cos^2\th\,\d\widetilde\phi_1 -
  \eta\sin^2\th\,\d\widetilde\phi_2\right).
\ee
There is a physical timelike singularity at $\bar{R}=0$, where the
Ricci scalar diverges, and a horizon
at $\bar{R}^2 = \bar{Q}$.  As we will observe below, the $\bar{\beta}$
dependence drops out of the metric at the horizon.

To calculate its area, consider the induced metric on a
$\bar{r}, \bar{t} ={\rm const}$ surface,
\begin{eqnarray}
\d s_\mathrm{ind}^2 & = & \mathbb{H}^{\frac{1}{3}}\bar{r}^2\left(\d\th^2 +
\cos^2\th\,\d\widetilde\phi_1^2 +
\sin^2\th\,\d\widetilde\phi_2^2\right) \nonumber \\ 
 & & \quad - \mathbb{H}^{-\frac{2}{3}} \left(\bar{\beta}\bar{r}^2 +
\bar{\beta}(\bar{Q}_1+\bar{Q}_2) + \frac{J_\beta}{2\bar{r}^2}\right)^2
\left(\cos^2\th\,\d\widetilde\phi_1 -
\eta\sin^2\th\,\d\widetilde\phi_2\right)^2.
\label{eqn:ind}
\end{eqnarray}
The determinant at $\bar{r}=0$ takes the simple form
\begin{equation}
\left. \phantom{\frac{1}{2}} h_\mathrm{ind} \right|_{\bar{r}=0} =
\left(\bar{Q}_1\bar{Q}_5\bar{Q}_{k\beta} -
\frac{J_\beta^2}{4}\right)\sin^2\th\cos^2\th,
\end{equation}
giving the horizon area
\begin{equation}
\mathcal{A} = \left. \phantom{\frac{1}{2}} \int_{S^3}
  \sqrt{h_\mathrm{ind}} \right|_{\bar{r}=0}=
  2\pi^2\sqrt{\bar{Q}_1 \bar{Q}_5 \bar{Q}_{k\beta}
  -\frac{J_{\beta}^2}{4}}.
\end{equation}
This result is appealing, since it takes the standard form of the
horizon area of a three--charge black hole, but in terms of the
charges which are all modified by the G\"odel parameter, $\beta$.
In terms of the original charges, however, it turns out that we can
equally well write the above horizon area as
\begin{equation}
\mathcal{A} = 2\pi^2\sqrt{{Q_1Q_5Q_k}-\frac{J^2}{4}},
\end{equation}
so this \emph{is} still an example of the conjecture
of~\cite{Gimon:ppblackstrings}, that such
a result should be independent of the G\"odel parameter.  In either
case, there is a bound on the
angular momentum which takes the same form as that for the
three--charge black hole:
\be
J^2 \le 4 Q_1Q_5Q_k \qquad \Leftrightarrow \qquad J_\beta^2 \le 4
\bar{Q}_1\bar{Q}_5\bar{Q}_{k\beta}.
\ee
The characterisation of~\cite{gibbons:99} as under-- or over--rotating is
equally applicable in the G\"{o}del background, in terms of the
modified parameters.

This characterisation is also relevant to the discussion of CTCs, which are
present for
\begin{equation}
\bar{f}(\bar{r}) \equiv \bar{r}^4 \left(\bar{\beta}\bar{r}^2 +
  \bar{\beta}(\bar{Q}_1+\bar{Q}_5) +
  \frac{J_\beta}{2\bar{r}^2}\right)^2 -\bar{r}^6\mathbb{H} > 0.
\end{equation}
The general condition for the existence of CTCs at a
radius $r$ in terms of the original parameters is therefore
\begin{equation} \label{eqn:CTCs}
f(r) \equiv \frac{J^2}{4} - (r^2+Q_1)(r^2+Q_5)(r^2+Q_k) + \beta^2 r^4
(r^2+Q_1)(r^2+Q_5) > 0.
\end{equation}
The rotating three--charge black hole is classified as over--rotating
when $J^2 > 4Q_1Q_5Q_k$~\cite{gibbons:99}, whence there are CTCs
induced by the rotation of the branes, outside of the horizon
itself\,\footnote{In the $\beta=0$ case, this corresponds to a breakdown
of unitarity in the dual field theory~\cite{Herdeiro:rotate}.}.
Correspondingly, for $J^2 < 4Q_1Q_5Q_k$, the under--rotating case
of~\cite{gibbons:99}, these CTCs are hidden entirely behind the
horizon.  Indeed, for $\beta=0$, the condition for CTCs reduces to that for the
three--charge black hole
\begin{equation}
J^2 > 4(r^2+Q_1)(r^2+Q_5)(r^2+Q_k).
\end{equation}
Thus CTCs exist outside the black hole horizon if $J^2 > 4Q_1Q_5Q_k$,
but these spacetimes are sick from varying
viewpoints~\cite{gibbons:99,Herdeiro:rotate,Jarv:CTC,Dyson:chronology}.
The rotation induced by the G\"{o}del background does not
change these properties, since the deformation vanishes as $r
\rightarrow 0$.  What the G\"{o}del rotation \emph{does} do, however,
is give rise to further CTCs at large $r$.  In the pure G\"{o}del
case, there are CTCs when $r > 1/\beta$.  For $r$ much larger than
any length scale associated with the black hole, they persist in the
more general system also.  (As $\beta \rightarrow 0$, these CTCs are
pushed off to infinity.)

There are other simple results which may be extracted from
(\ref{eqn:CTCs}) .  For a non--rotating
black hole ($J=0$) in a G\"odel universe, the condition reduces to
\be
r^2 > \frac{1}{\beta^2} + 2 Q_k.
\ee
Relative to the G\"{o}del background, the presence of the black hole
charge takes the VL surface, which we shall denote by $r_*$, to a
larger radius.

In order to consider more general results regarding the existence of
CTCs in the G\"odel black hole, we should study the
equation~(\ref{eqn:CTCs}) without setting any of the parameters to
zero.  Before doing so, however, we will consider the notion of
``holographic protection of chronology''.  In~\cite{Boyda:goedel}, it was
demonstrated that, relative to an observer at $r=0$, there is a
preferred holographic
screen in the G\"odel universe located at $r=r_\mathcal{H} \equiv
\sqrt{3}/(2\beta)$.  Since $r_\mathcal{H} < r_*$, there are no CTCs
enclosed within the screen, and in this sense chronology is protected;
the holographic screen carves out a causally well--behaved region of
the spacetime.  It is of interest to investigate if this
possible method of protecting chronology can exist in the G\"odel black
hole metric constructed here.  As in~\cite{Boyda:goedel}, the
spherical symmetry of the metric
allows us to construct the preferred holographic screen with relative
ease.  It is simply the constant $r$ surface of maximal area (with respect to
$r$).  From (\ref{eqn:ind}), we have
\be
\mathcal{A}(r)  =  \int_{S^3}\sqrt{h_\mathrm{ind}} =
2\pi^2 \sqrt{-f(r)}.
\ee
Maximising $\mathcal{A}(r)$ involves extracting the roots of
\begin{equation}
\frac{\del}{\del r} f(r) = 0,
\end{equation}
which is explicitly written as
\begin{equation}
r\left((Q_1+Q_5+2r^2)(Q_k+r^2-\beta^2r^4) +
(Q_1+r^2)(Q_5+r^2)(1-2\beta^2r^2)\right) = 0.
\end{equation}
We can observe immediately that the position of any holographic
screen is independent of $J$.  Furthermore, there is always a screen at
$r=0$, associated with the black hole horizon.  The
second screen, which was interpreted as a chronology protecting
screen in the G\"odel universe~\cite{Boyda:goedel}, only exists when
$\beta \neq 0$ as expected.  In the case in which we switch off the three
black hole charges, $Q_1$, $Q_5$ and $Q_k$, this screen occurs at
$r=\sqrt{3}/(2\beta)$, as in~\cite{Boyda:goedel}.  However, the fact
that the result is independent of $J$, whilst $J$ does affect the
position of the VL surface, raises the question of whether the screen
always lies within the VL surface.  We proceed to study this issue by
investigating the function $f(r)$.

Since $f$ is a function of $r^2$ only, we simplify this study by
considering $f$ as a quartic function in $x=r^2$:
\begin{equation}
f(x) = \frac{J^2}{4} - (x+Q_1)(x+Q_5)(x+Q_k) +
\beta^2x^2(x+Q_1)(x+Q_5).
\end{equation}
Note that $x=0$ represents the horizon of the black hole and the
physical values of $x$ are given by the range $x>0$.  Within this
range, a
positive value of $f$ will imply the existence of CTCs and an
extrema of $f$ will represent a holographic screen.  The general
behaviour of $f$ can be constructed by considering first
$f(x;\,J=\beta=0)$.  This function has leading behaviour $f \sim -x^3$
as $\vert x\vert \to \infty$, and three roots, each at negative values
of $x$.  In particular, we thus know that $f'(0;\,J=\beta=0) < 0$ and,
for $x \ge 0$, $f(x;\,J=\beta=0)$ is a monotonically decreasing
function.

Now consider the effect of the $J^2$--term on this function,
\emph{i.e.}, consider $f(x;\,\beta=0)$.  Given that this additional term is
independent of $x$, its effect is just to translate the previous
graph up by some constant amount.  The value of $f(x;\,\beta=0)$ at
$x=0$ is determined by the
size of $J$ relative to the product of the charges, $Q_i$, and this
determines if CTCs associated with the black hole are present outside
of the horizon, {\it i.e.}, whether we are considering an
over-- or under--rotating black hole.  Consider for now the
under--rotating case, for which we recall that $f(0;\,\beta=0) < 0$,
$f'(0;\,\beta=0) < 0$ and $f(x;\,\beta=0)$ is monotonically decreasing
for $x \ge 0$.

\begin{figure}[ht]  
\begin{center} 
\includegraphics[width=0.5\textwidth, height=0.25\textheight]{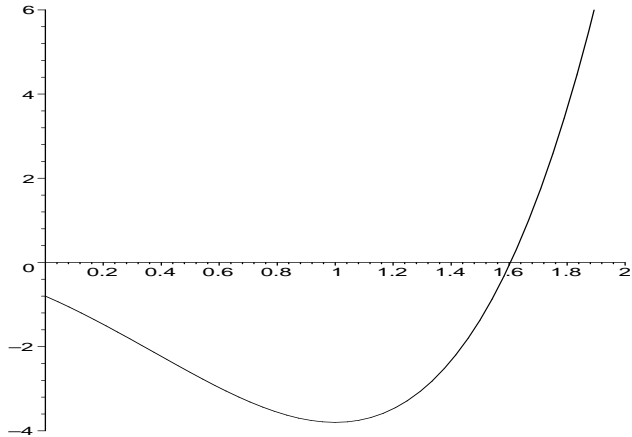} 
\end{center} 
\caption{Characteristic behaviour of $f(x)$ for the under--rotating
  black hole, plotted in the dimensionless variables, $\beta^6 f$ {\it
  vs.} $\beta^2 x$.  CTCs occur where $f(x)>0$ and the holographic
  screen is at the minimum.}
\label{fig:CTC1}
\end{figure}

The final term which we add to this function is the $\beta^2$--term
which is monotonically increasing in the range of interest, $x \ge 0$.
This term has the important properties that $f(0;\, J=Q_i=0)=0$ and
$f'(0;\, J=Q_i=0)=0$, whilst its leading behaviour, $f\to\beta^2 x^4$
as $x \to \infty$, dominates that of the existing cubic function.
Therefore the characteristic behaviour of the complete quartic
function, $f(x)$, in the region of interest is always of the form
given in figure~\ref{fig:CTC1}.  We note in particular that, despite
the presence of the rotating black hole, there always exists a
holographic screen, outside of the black hole horizon, which shields
the CTCs associated with the G\"odel--like asymptotics.  Furthermore,
the single minimum of the function ensures that the mixing of the
G\"odel rotation and black hole rotation never generates extra
isolated regions of CTCs, other than those which first occur at the VL
surface, $r=r_*$, and persist all the way out to infinity --- recall,
however, that the position of this VL surface {\it is} dependent on
the black hole parameters.

\begin{figure}[ht]  
\begin{center} 
\includegraphics[width=0.5\textwidth, height=0.25\textheight]{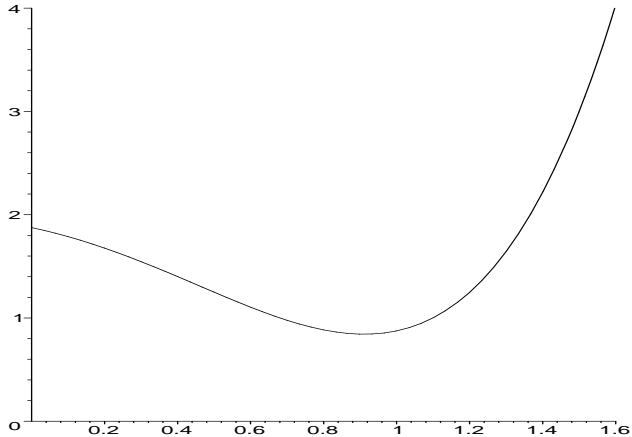} 
\end{center} 
\caption{Characteristic behaviour of $f(x)$ for the over--rotating
  black hole with CTCs everywhere (which arises for sufficiently large
  $\beta$), plotted in the dimensionless variables, $\beta^6 f$ {\it
  vs.} $\beta^2 x$.}
\label{fig:CTC2}
\end{figure}

For completeness, consider what occurs in the case of the
over--rotating G\"odel black hole.  The arguments of the
previous paragraphs remain, but we note that there is no reason for the
CTCs associated with the over--rotating black hole not to mix with
those associated with the G\"odel universe.  Indeed, if the bound on
the black hole angular momentum, $J \le 4 Q_1 Q_5 Q_k$, is violated,
then for sufficiently large values of the G\"odel parameter, $\beta$,
the minimum in $f(x)$ will occur in the region where the black hole
CTCs still persist --- the spacetime outside of the horizon will then
be causally sick everywhere.  The behaviour of $f(x)$ in this case is
illustrated in figure~\ref{fig:CTC2}.

It is important that we have been able to draw these conclusions only
because we have a proper (ten--dimensional) microscopic understanding
of the G\"odel black hole.
Consider, that is, asking similar questions regarding the possibility of
holographic protection of chronology in the under--rotating
BMPV--G\"odel black hole of~\cite{Herdeiro:BMPVgoedel}.  This arises
from setting $\bar{Q}_1 = \bar{Q}_5 = \bar{Q}_{k\beta} = \bar{Q}$
in~(\ref{eqn:5D3charge}).  In that case, one can follow the line of
argument described previously, but now beginning with the function
\begin{equation}
\bar{f}(\bar{x}) = \frac{J_\beta^2}{4} - (\bar{x}+\bar{Q})^3 +
\bar{\beta}(x^2+2\bar{Q}\bar{x})
\left(\bar{\beta}(x^2+2\bar{Q}\bar{x})+J_\beta\right).
\end{equation}
The difference arises in the fact that the $\bar{\beta}$--term does not
have a vanishing first derivative.  It is thus possible that, for a
given choice of parameters, the minimum in $\bar{x}>0$ does not occur
and one would conclude that holographic protection of chronology
fails.  This indeed occurs when we pick parameters satisfying
\begin{equation}
16\bar{\beta}^2 \left(2\bar{\beta}J_\beta - 3\bar{Q}\right) >
\left(3-8\bar{\beta}^2 \bar{Q}\right)^2,
\end{equation}
a condition which arises by solving explicitly for the roots of
$\bar{f}'(\bar{x}) = 0$.

Without an underlying microscopic knowledge
of the parameters, such an inequality can be satisfied.  Consider,
however, the subsequent requirement that $2\bar{\beta}J_\beta -
3\bar{Q} > 0$.  Converting back to the parameters, $J$, $\beta$ and $Q$,
this inequality is equivalent to
\begin{equation}
J -\frac{1}{2\beta}\left(3Q+2\beta^2Q^2\right) > 0.
\end{equation}
Given that, for an under--rotating black hole we have $J^2 < 4Q^3$, we
know that this can only be satisfied if
\begin{equation}
16\beta^2Q^3 - \left(3Q+2\beta^2Q^2\right)^2 = -4Q^2\left( \left( \beta^2
Q - \frac{1}{2} \right)^2 + 2 \right) > 0,
\end{equation}
an obvious impossibility --- the
parameters are not allowed to be chosen in such a way that would
violate the holographic protection of chronology!  The phase space of
the parameters $\bar{Q}$, $\bar{\beta}$ and $J_{\beta}$ is thus
reduced by our knowledge of the underlying microscopic description of the
BMPV G\"odel black hole.

\sect{Discussion} \label{sect:conc}

In this paper we have found a solution corresponding to a rotating,
charged black hole in a G\"odel universe.  Our solution generalises
previously known solutions and provides a basis for a microscopic
description in terms of D-branes.  We hope that it will serve
as a laboratory to investigate the issue of CTCs in string theory.  In
particular, we hope that it can throw light on possible mechanisms that
either prevent the formation of CTCs, or else render them harmless.

It has been argued~\cite{Dyson:chronology}, for example, that
there is just such a mechanism at work in the case of rotating, charged
black holes in an asymptotically flat background.  When one attempts
to build such a black hole step--by--step, by throwing in matter, it
is found that the total angular momentum can never exceed the bound
for which CTCs are formed.  The creation of a black hole with CTCs is
therefore claimed to be impossible~\cite{Dyson:chronology}.  Similar
arguments concerning the G\"odel universe have been advanced
in~\cite{drukker:03}.

For such G\"odel universes, it has been argued~\cite{Boyda:goedel} that
there are holographic screens which shield local
observers from the effects of any (G\"odel) CTC.  In the solution that
we have found, one can study both types of CTCs simultaneously and it
would be very interesting to analyse in what sense the two classes of
CTCs differ and in what sense they are similar.

It is by no means obvious that the existence of the holographic
screens guarantee that the G\"odel CTCs are harmless.  In fact, as
shown in~\cite{Hikida:goedel}, it seems quite possible for a probe to
follow a CTC all the way round -- despite the presence of the
holographic screens -- and it is not at all clear how paradoxes can be
avoided.  An alternative possibility~\cite{drukker:03}, more in line
with~\cite{Dyson:chronology}, is simply to say that a G\"odel universe
can not be formed in the first place and that the solution therefore
is unphysical.  In~\cite{Dyson:chronology} the argument was based on
whether local physics could give rise to CTCs in a universe that was
well behaved on large scales.  In case of the G\"odel universe -- with
CTCs everywhere -- it is not clear how a similar argument can be
formulated in any interesting way.  Clearly much remains to be understood.

\vskip .5in

\centerline{\bf Acknowledgments}

DB would like to thank Carlos Herdeiro for some useful correspondence,
and Paul Saffin and Ehud Schreiber for discussions.  JPG would like to
thank Neil Constable for early discussions which prompted the start of
this project.  DB is supported in part by NSERC, UD is a Royal Swedish
Academy of Sciences Research Fellow supported by a grant from the Knut
and Alice Wallenberg Foundation and this work was also supported by
the Swedish Research Council (VR).

\medskip


\appendix

\renewcommand{\theequation}{\Alph{section}.\arabic{equation}}

\sect{Appendix: Ten--dimensional supergravity theories}
\label{sect:eom}

A consistent truncation of the bosonic sector of type
IIB supergravity is provided by the metric, $g_{ab}$, the dilaton,
$\Phi$, and the Ramond--Ramond (RR) potentials, $C_2$ and $C_4$.
The RR potentials give rise to the gauge invariant field strengths,
$F_3=\d C_2$ and $F_5=\d C_4$, the latter being
self--dual.  One can, however, take the action to be~\cite{bergshoeff:96}
\be
2\ka_{10}^2  \,S_{\rm IIB} = \int \d^{10} x
  \sqrt{-g_{10}} \left( R_{10} -
  \frac{1}{2} \del \P \cdot \del \P \right) - \frac{1}{2} \int
\left( e^{\P} \star F_3 \wedge F_3 + \frac{1}{2} \star F_5 \wedge
  F_5 \right) ,
\ee

\noindent and impose the self--duality condition, $\star F_5 =
F_5$, at the level of the equations of motion, which are~\cite{schwarz:83}
\begin{eqnarray}
&& \mathcal{R}_{ab} = \frac{1}{2}\del_a\Phi\del_b\Phi +
\frac{1}{96}F_{ac_1\ldots c_4}F_{b}^{~~c_1\ldots c_4} +
\frac{1}{4}e^\Phi\left(F_{acd}F_{b}^{~~cd} -
  \frac{1}{12}g_{ab}F_3^2\right), \nonumber \\
&& \Box\Phi = \frac{1}{12}e^\Phi F_3^2, \qquad \d F_5 = 0, \qquad F_5
= \star F_5, \label{eqn:iib} \\
&& \d\left(e^\Phi \star F_3\right) = 0, \qquad F_5 \wedge F_3 = 0. \nonumber
\end{eqnarray}

The complete set of T--duality rules to take us to the IIA theory can
be found in, \emph{e.g.},~\cite{Johnson:dbranes}.  In our case,
T--duality along $y_5$ gives the type IIA fields
\begin{eqnarray}
\tilde{g}_{5 5} & = & \frac{1}{g_{5 5}}, \qquad \quad
\tilde{g}_{a b} = g_{a b} -\frac{g_{a 5}g_{b 5}}{g_{5 5}}, \qquad
\tilde{B}_{a 5} =  \frac{g_{a 5}}{g_{5 5}}, \qquad
e^{2\tilde\Phi} =  \frac{e^{2\Phi}}{g_{5 5}}, \nonumber\\
\tilde{C}^{(1)}_a & = & C^{(2)}_{5 a}, \qquad \tilde{C}^{(3)}_{a b c}
=  C^{(4)}_{5 a b c}, \qquad \tilde{C}^{(3)}_{5 a b} = C^{(2)}_{a b} -
\frac{2}{g_{5 5}} g_{5  [a}C^{(2)}_{\vert 5 \vert b]},
\label{eqn:tdual} \\
\tilde{C}^{(5)}_{5 a b c d} & = & C^{(4)}_{a b c d} - \frac{4}{g_{5 5}}
  g_{5[a}C^{(4)}_{\vert 5 \vert b c d]},\nonumber
\end{eqnarray}

\noindent where $a,b$ run over all directions except $y_5$.  Note that
the T--duality here is acting on the string frame
metric, whereas everything in the text is written in terms of the
Einstein frame metric $g_{ab}^{\textrm{E}} =
e^{-\Phi/2}g_{ab}^{\textrm{S}}$.

The bosonic sector of type IIA supergravity is provided by the metric,
$g_{ab}$, the dilaton, $\Phi$, the Neveu Schwarz--Neveu Schwarz (NS--NS)
two--form potential, $B_2$, and the RR potentials, $C_1$
and $C_3$.  The potentials give rise to a NS--NS field strength,
$H_3 = \d B_2$, and the gauge invariant RR field strengths, $F_2 =
\d C_1$ and $G_4 = \d C_3 + C_1 \wedge H_3$.  The action is
\[
 2\ka_{10}^2  \, S_{\rm IIA} = \int \d^{10} x \sqrt{-g}
   \left( R -\frac{1}{2}  \del \P \cdot \del \P \right)
\]
\be
- \frac{1}{2} \int \left( e^{3\P/2} \star F_2
\wedge F_2 + e^{\P/2} \star G_4 \wedge G_4 + e^{-\P} \star
H_3 \wedge H_3 - B_2 \wedge \d C_3 \wedge \d C_3 \right),
\ee

\noindent with the equations of motion
\begin{eqnarray}
& & R_{ab} = \frac{1}{2} \del_a \Phi \del_b \Phi + \frac{1}{4} e^{-\Phi}
\left( H_{a cd} H_b^{~cd} - \frac{1}{12} g_{ab} H_3^2 \right)  \nonumber \\
&& ~~~~ + \frac{1}{2} e^{3\Phi/2} \left( F_{a c} F_b^{~c} -
\frac{1}{16} g_{ab} F_2^2 \right) + \frac{1}{4} e^{\Phi/2} \left( G_{a
  cde} G_b^{~cde} - \frac{3}{32} g_{ab} G_4^2 \right), \nonumber\\
& & \Box \Phi = -\frac{1}{12}e^{-\Phi}H_3^2 + \frac{3}{8}e^{3\Phi/2}F_2^2 +
\frac{1}{96}e^{\Phi/2}G_4^2, \\
&& \d\left(e^{3\Phi/2} \star F_2\right) = \,e^{\Phi/2} \star
G_4 \wedge H_3, \qquad \d\left(e^{\Phi/2} \star G_4\right) =
\,G_4 \wedge H_3, \nonumber\\
&& \d\left(e^{-\Phi} \star H_3\right) = -\,e^{\Phi/2} \star
G_4 \wedge F_2 + \frac{1}{2} G_4 \wedge G_4.\nonumber
\end{eqnarray}
In the text, we also have a five--form RR potential, $C_5$.  The
associated gauge invariant field strength is dual to $G_4$:
\be
G_6 = F_6 + C_3 \wedge H_3 = e^{\Phi/2} \star G_4, \qquad F_6 = \d C_5,
\ee
so the above equation of motion for $G_4$ becomes a Bianchi identity,
\be
\d G_6 = G_4 \wedge H_3,
\ee
for $G_6$.  The Bianchi identity, $\d G_4 = F_2 \wedge H_3$, for $G_4$
then becomes the equation of motion,
\be
\d \left( e^{-\Phi/2} \star G_6 \right) = - F_2 \wedge H_3,
\ee
for $G_6$.

\sect{Appendix: Five--dimensional supergravity theories}

The relevant reduction of type IIB supergravity to five dimensions is derived
in~\cite{herdeiro:02}.  We extend this to include the four--form RR
potential.  Reducing on $T^4 \times S^1$, the ansatz with
the five--dimensional metric written in
the Einstein frame is
\be
\d s^2_{10} = e^{2a\la} \left( e^{2 b\psi} \d s^2_5 + e^{-8a\la} ( \d
  y_5 + A_1 )^2 \right) + e^{-3b\psi /2} \d s^2(T^4),
\ee

\noindent Replacing $\la$ with\footnote{This corrects a typographical error
  in~\cite{herdeiro:02}.}
\be
\chi(x) = -\frac{1}{4} \frac{b}{a} \psi(x) - \la(x),
\ee

\noindent and taking $a^2=1/24$ and $b^2=1/9$, gives rise to canonically
 normalised five--dimensional scalars.  The dilaton reduces trivially
 and the two--form gauge field as
\be
C_2 = B_2 + C_1 \wedge \d y_5,
\ee

\noindent where $B_2$ and $C_1$ are five--dimensional gauge
potentials.  The modified field strength which appears in the
five--dimensional action is
\be
G_3 = H_3 - F_2 \wedge A_1, \qquad H_3 = \d B_2, \qquad F_2 = \d C_1.
\ee

\noindent For the four--form gauge field we take an ansatz adapted to
the specific case considered in the text, writing
\be
C_4 = \frac{1}{2} \d y_i \wedge \d y_j \wedge \left( \xi_2^{ij} +
  \xi_1^{ij} \wedge \d y_5 \right).
\ee

\noindent The notation $\xi_p^{ij} = -\xi_p^{ji}$ denotes a collection of
five--dimensional $p$--forms.  With
$F_3^{ij} = \d \xi_2^{ij}$ and $F_2^{ij} = \d \xi_1^{ij}$, the
five--form field strength is
\be
F_5 = \frac{1}{2} \d y_i \wedge \d y_j \wedge \left( F_3^{ij} +
  F_2^{ij} \wedge \d y_5 \right),
\ee

\noindent whereas the natural gauge invariant object is
\be
G_3^{ij} = F_3^{ij} - F_2^{ij} \wedge A_1.
\ee

Defining
\be
\Phi_\pm = 2 \left( \pm 2a\chi - \frac{3b}{2} \psi \right),
\ee

\noindent the five--dimensional action becomes~\cite{herdeiro:02}
\[
2\ka_5^2 \,S_5 = \int \d^5 x\sqrt{-g_5} \left( R_5 -
  \frac{1}{2} \del \P \cdot \del \P - \frac{1}{2}\del \psi \cdot \del
  \psi - \frac{1}{2}\del \chi \cdot \del
  \chi \right)
\]
\be
- \frac{1}{2} \int \left( e^{8a\chi} \star F_{(1)} \wedge
    F_{(1)} + e^{\P + \P_+} \star G_3 \wedge G_3 +
    e^{\P + \P_-} \star F_{(2)} \wedge F_{(2)} \right)
\ee
\[
-\frac{1}{2} \int  \sum_{j>i} \left( e^{4a\chi}
        \star G^{ij}_3 \wedge G^{ij}_3 +
    e^{-4a\chi} \star F^{ij}_2 \wedge F^{ij}_2 \right),
\]

\noindent where $F_{(1)} = \d A_1$ is the two--form arising from the metric
and $F_{(2)} = \d C_1$ is the two--form from the RR field.  The
Einstein equation is
\ba
&&R_{ab} = \frac{1}{2} \left( \del_a \P \del_b \P + \del_a \psi \del_b
  \psi + \del_a \chi \del_b \chi \right) + \frac{1}{2} e^{8a\chi}
\left( F_{(1)ac}F^{~~~c}_{(1) b} - \frac{1}{6}
  F^2_{(1)} g_{ab} \right) \nonumber\\
&& \qquad + \frac{1}{4} e^{\P+\P_+} \left(
  G_{acd}G_b^{~cd} - \frac{2}{9} G^2_3 g_{ab} \right) +
\frac{1}{2} e^{\P+\P_-} \left(
  F_{(2)ac}F^{~~~c}_{(2)b} - \frac{1}{6} F^2_{(2)} g_{ab} \right) \\
&& \qquad \qquad +  \sum_{j>i} \left[
  \frac{1}{4} e^{4a\chi} \left( G^{ij}_{acd}G^{ij\,cd}_b - \frac{2}{9}
    (G^{ij}_3)^2 g_{ab} \right) +
\frac{1}{2} e^{-4a\chi} \left(
  F^{ij}_{ac}F^{ij\,c}_{b} - \frac{1}{6} (F^{ij}_2)^2 g_{ab} \right)
\right], \nonumber
\label{eqn:5dEin}
\ea
and the scalar equations are
\ba
\Box \P &=& \frac{1}{12} e^{\P+\P_+} G_3^2 +
\frac{1}{4}e^{\P+\P_-} F_{(2)}^2 = -\frac{1}{3b} \Box \psi,
\label{eqn:5d} \\
\Box \chi &=& \frac{a}{3} \left( e^{\P+\P_+} G_3^2 +\sum_{j>i}
  e^{4a\chi} (G_3^{ij})^3 \right) -
a \left( e^{\P+\P_-} F_{(2)}^2 + \sum_{j>i} e^{-4a\chi} (F_2^{ij})^2
  \right) \nonumber\\
&& \qquad + 2a e^{8a\chi} F_{(1)}^2, 
\ea
which allow for the solution $\psi = -3b \P$.  The form field equations are
\ba
&&\d (e^{8a\chi} \star F_{(1)} ) = e^{\P+\P_+} \star G_3
\wedge F_{(2)} +  \sum_{j>i}
e^{4a\chi} \star G^{ij}_3 \wedge F^{ij}_2, \\
&& \d (e^{\P+\P_+} \star G_3) = 0, \qquad \d (e^{4a\chi} \star
G^{ij}_3) = 0, \label{eqn:G} \\
&& \d \left( e^{\P+\P_-} \star F_{(2)} \right) = e^{\P+\P_+} \star G_3
\wedge F_{(1)}, \qquad \d \left( e^{-4a\chi} \star F^{ij}_2 \right) =
  e^{4a\chi} \star G_3^{ij} \wedge F_{(1)},
\ea

\noindent and the $G_3$ equations (\ref{eqn:G}) can be solved by
  dualising as
\be
G_3 = e^{-(\P + \P_+)} \star \d \hat{A}_1 \equiv e^{-(\P + \P_+)}
\star \hat{F}_2, \qquad G^{ij}_3 = e^{-4
a\chi} \star \d \hat{A}^{ij}_1 \equiv e^{-4
a\chi} \star \hat{F}_2^{ij}.
\ee
There are also the non--trivial Bianchi identities
\be
\d G_3 = -F_{(2)} \wedge F_{(1)}, \qquad \d G^{ij}_3 = -F^{ij}_2 \wedge
F_{(1)}.
\label{identity}
\ee

The minimal supergravity theory~\cite{cremmer} in five dimensions has
no scalars and a single gauge field.  Up to rescalings, we
recover~\cite{herdeiro:02} the equations of motion of that
theory,
\be
\d \star F + F \wedge F = 0,
\label{eqn:min}
\ee
by taking $\P = \psi = \chi = 0$ in the above and setting all the
gauge fields to be equal.  Since
we will only be interested in obtaining the minimal theory in one of
two cases in the text, we first consider $F_2^{ij}=G_3^{ij}=0$
(relevant to the BMPV black hole) and
\be
F_{(1)}=a_1 F, \qquad F_{(2)}=a_2 F, \qquad G_3 = a_1 a_2 \star F,
\ee
where $a_1^2=a_2^2=1$.  Then the form equations (\ref{eqn:G}) and
Bianchi identity (\ref{identity}) all give the minimal equation
(\ref{eqn:min}).  We can also
consider $F_{(2)} = G_3 = 0$ (relevant to the G\"{o}del universe) and
\be
F_{(1)} = aF, \qquad F_2^{ij} = c^{ij} F, \qquad G_3^{ij} = 
ac^{ij} \star F,
\ee
with
\be
a^2 = 1, \qquad \sum c^{ij} c^{ij} = 1,
\ee
which also gives the minimal equation (\ref{eqn:min}).

Both the BMPV and G\"{o}del solutions can be written as
\ba
&&\d s^2 = -H^{-2}(\d t - \s )^2 + H \d s^2 (\mathbb{R}^4), \\
&&A = H^{-1} ( \d t - \s ) - \d t, \label{eqn:A}
\ea
and we define $J = \d \s$.  Then the equation (\ref{eqn:min}) is
solved for harmonic $H$ and
\be
\star_4 J = -J.
\ee
Now take
\be
\d s^2(\mathbb{R}^4) = \d r^2 + r^2 \left( \d \th^2 + \cos^2 \th \d
  \p_1^2 + \sin^2 \th \d \p_2^2 \right),
\ee
and
\be
\s = f(r) \left( \cos^2 \th \d \p_1 - \eta \sin^2 \th \d \p_2 \right).
\ee
We have $\star_4 J = -J$ for 
\be
f(r) = c r^{-2\eta}.
\ee
So we have the BMPV black hole for $\eta = +1$ and the
G\"{o}del universe for $\eta = -1$.  We should note that this does not
appear to agree with the usual
conventions~\cite{gauntlett,Herdeiro:rotate}, in which the G\"{o}del
universe has non--zero $J_R$ ($\eta=+1$) and the BMPV black
hole has non--zero $J_L$ ($\eta=-1$).  To agree with this, we would
need to reduce the negative charge sector of the IIB theory.  Indeed, if we
flip the sign of the gauge field in (\ref{eqn:A}) above, we can
recover the other
two solutions (the BMPV black hole for $\eta = -1$ and the
G\"{o}del universe for $\eta = +1$).  However, we would also need to take
an \emph{anti}--self--dual five--form in ten dimensions, which is why
we have decided to
take the ``opposite'' conventions to those
of~\cite{gauntlett,Herdeiro:rotate}.

\noindent

\medskip



\providecommand{\href}[2]{#2}\begingroup\raggedright\endgroup

\end{document}